%% file: main.tex
\documentclass[acmsmall,screen,nonacm]{acmart}

\usepackage{textcomp}
\usepackage{tabularx}
\usepackage{array}
\usepackage{multirow}
\usepackage{adjustbox}
\usepackage{xurl}
\usepackage[most]{tcolorbox}
\usepackage{subcaption}
\DeclareUrlCommand{\code}{}

\newtcolorbox{findingbox}{
  enhanced,
  breakable,
  colback=gray!8,
  colframe=gray!45,
  boxrule=0.6pt,
  arc=2mm,
  left=1.0em,
  right=1.0em,
  top=0.7em,
  bottom=0.7em,
  before skip=0.9em,
  after skip=0.9em
}
\newcommand{\finding}[2]{%
  \begin{findingbox}
    \noindent\textbf{\emph{Finding #1:}} \emph{#2}
  \end{findingbox}%
}

\newtcolorbox{answerbox}{
  enhanced,
  breakable,
  colback=gray!10,
  colframe=gray!10,
  boxrule=0pt,
  arc=0pt,
  left=1.0em,
  right=1.0em,
  top=0.8em,
  bottom=0.8em,
  borderline west={3pt}{0pt}{gray!60},
  before skip=1.0em,
  after skip=1.0em
}

\setcopyright{none}
\renewcommand\footnotetextcopyrightpermission[1]{}

\begin{document}

\title{Understanding Verification Vulnerabilities in Cross-chain Bridges: An Empirical Study}

\author{Yuying Wang}
\affiliation{%
  \institution{Institute of Computing Technology, Chinese Academy of Sciences}
  \city{Beijing}
  \country{China}
}
\affiliation{%
  \institution{University of Chinese Academy of Sciences}
  \city{Beijing}
  \country{China}
}

\author{Yi Guo}
\affiliation{%
  \institution{Institute of Computing Technology, Chinese Academy of Sciences}
  \city{Beijing}
  \country{China}
}

\author{Jiachi Chen}
\affiliation{%
  \institution{The State Key Laboratory of Blockchain and Data Security, Zhejiang University}
  \city{Hangzhou}
  \country{China}
}
\affiliation{%
  \institution{Hangzhou High-Tech Zone (Binjiang) Institute of Blockchain and Data Security}
  \city{Hangzhou}
  \country{China}
}

\author{Hanwen Zhang}
\affiliation{%
  \institution{Institute of Computing Technology, Chinese Academy of Sciences}
  \city{Beijing}
  \country{China}
}
\affiliation{%
  \institution{University of Chinese Academy of Sciences}
  \city{Beijing}
  \country{China}
}

\author{Yi Sun}
\affiliation{%
  \institution{Institute of Computing Technology, Chinese Academy of Sciences}
  \city{Beijing}
  \country{China}
}
\affiliation{%
  \institution{University of Chinese Academy of Sciences}
  \city{Beijing}
  \country{China}
}

\renewcommand{\shortauthors}{Wang et al.}

\begin{abstract}
With the rapid development of blockchain technology, numerous heterogeneous blockchains have emerged, each featuring distinct architectures, consensus mechanisms, and application ecosystems. Cross-chain bridges have consequently become a critical infrastructure component, typically relying on smart contracts to facilitate message passing and asset transfers across heterogeneous blockchains. Unlike single-chain contracts, bridge contracts introduce additional verification logic to validate external-chain data, such as state proofs, message sender identities, and signatures. However, correctly implementing such logic is challenging. We refer to vulnerabilities that violate the security guarantees of cross-chain verification as Cross-chain Verification Vulnerabilities (CVVs). To characterize CVVs, we conduct the \textit{first} systematic empirical study of this vulnerability class. By analyzing 588 security audit reports from 36 real-world cross-chain bridge projects, we derive 11 CVV patterns and organize them into 5 high-level categories according to the affected verification properties. Following this taxonomy, we construct a CVV dataset containing 830 vulnerability instances across 786 contract files. The dataset establishes mappings among audit reports, audited source-code versions, affected source files, and CVV-pattern labels. We then assess existing detection approaches. Current tools provide limited documented coverage of CVV patterns, while only two of four evaluated LLMs outperform an all-pattern constant baseline of 29.3\% Macro-F1; the best reaches 51.2\%. Together, the limited tool coverage and inconsistent LLM gains motivate detection methods tailored to cross-chain verification semantics.
\end{abstract}



\maketitle

\input{section-01-Introduction}

\input{section-02-Background}
\input{section-03-vulnerability-taxonomy}
\input{section-04-vulnerability-dataset}
\input{section-05-vulnerability-detection-evaluation}

\input{section-06-Discussion}
\input{section-07-related-work}
\input{section-08-Conclusion}

\section{Data Availability}
A replication package containing the dataset and evaluation results is available at \url{https://anonymous.4open.science/r/CVV-Dataset/}.

\bibliographystyle{ACM-Reference-Format}
\bibliography{ref}

\end{document}

%% file: section-01-Introduction.tex
\section{Introduction}

Smart contracts have become critical software components for managing digital assets and enforcing application logic on blockchains~\cite{smart-contracts-security}. As multi-chain ecosystems expand, smart contracts increasingly coordinate cross-chain operations in bridge protocols~\cite{cross-chain-bridge-1,cross-chain-bridge-2}. Unlike single-chain contracts, bridge contracts must verify the validity and trustworthiness of cross-chain messages, state proofs, signatures, and other data from external chains~\cite{bridge-Verification-1}. Cross-chain verification is implemented via diverse mechanisms~\cite{verify-mechanism}, such as multisignature schemes~\cite{layerzero,wormhole,chainlink}, light clients~\cite{ibc,rainbow,ETH-Relay}, and zero-knowledge proofs~\cite{zkBridge,zkRelay}. Their distinct trust assumptions and security boundaries make cross-chain verification logic complex and prone to implementation vulnerabilities.

Recent cross-chain bridge incidents further demonstrate the severity of cross-chain verification vulnerabilities. As of June 2026, we identified at least 11 bridge attacks involving such vulnerabilities, resulting in about \$330 million in asset losses~\cite{slowmist_bridge_hack_db}. For example, the Kelp DAO rsETH bridge attack caused an estimated \$292 million loss through forged LayerZero cross-chain messages~\cite{kelp}, while the Hyperbridge attack used forged state proofs to gain unauthorized privileges and mint one billion wDOT~\cite{hyperbridge}. Although these incidents targeted different bridge designs, they share a common failure mode: invalid external-chain data was trusted and used to authorize cross-chain operations. 


In this paper, we define \emph{Cross-chain Verification Vulnerabilities (CVVs)} as vulnerabilities that undermine the security guarantees of cross-chain verification. CVVs arise when bridge contracts incorrectly authenticate, interpret, or process external-chain data, such as messages, proofs, or signatures. For example, an access-control flaw is a CVV only if it causes data from an unauthorized cross-chain sender to be accepted.


Existing smart contract security research has largely focused on single-chain vulnerabilities~\cite{vulnerability-survey-1,vulnerability-survey-2}, typically classifying them by implementation-level defects, such as access-control errors~\cite{AChecker,access-control-2} and integer overflows~\cite{integer-overflow-1,integer-overflow-2}. While useful for describing how an implementation fails, such classifications do not capture which cross-chain verification guarantee is violated in verification-critical contexts. Recent studies on cross-chain bridge security~\cite{smartaxe,bridgeguard,Xscope} mainly target specific scenarios, vulnerability patterns, or detection rules, leaving cross-chain verification vulnerabilities without a systematic semantic characterization. Meanwhile, although LLMs have shown promise in smart contract vulnerability detection~\cite{LLM-detect-1,LLM-detect-2}, their ability to reason about these verification-semantic vulnerabilities remains unclear.


To fill this gap, we conduct, to the best of our knowledge, the first systematic empirical study of cross-chain verification vulnerabilities. 
We collect and analyze 588 security audit reports from 36 real-world cross-chain bridge projects~\cite{chainspot_bridge_db,defillama_bridge_tracker}. 
We identify CVV instances from the collected audit reports based on a set of criteria.
Based on these instances, we use open card sorting~\cite{card-sorting} and incorporate cross-chain verification domain knowledge to derive CVV patterns. 
We identify 11 CVV patterns and organize them into 5 high-level categories according to the cross-chain verification properties they violate (e.g., message authenticity and execution integrity).

Based on this taxonomy, we construct a real-world CVV dataset. For each vulnerability instance, we trace the corresponding audit report to its project repository and audited source-code version. We then identify the affected source files and map the instance to a specific CVV pattern. The resulting dataset contains 830 vulnerability instances across 786 contract files. Using this dataset, we characterize CVVs from multiple perspectives, including vulnerability distribution, severity, and cross-project diversity. These analyses reveal several key findings on how CVVs manifest within and affect real-world cross-chain bridges.

Finally, we evaluate the applicability of existing methods to CVV detection, including general smart contract detectors, bridge-oriented detectors, and mainstream LLMs. General-purpose tools mainly target implementation-level defects, while bridge-oriented approaches cover only selected scenarios or vulnerability patterns. LLMs are more flexible, but their performance remains inconsistent, with only two of the four evaluated models outperforming a simple constant baseline. Error analysis further attributes most failures to incomplete cross-component context and limited cross-chain semantic reasoning. These results motivate CVV-specific detection techniques that better capture cross-component context and verification semantics.

We summarize our main contributions as follows:

\begin{itemize}
    \item We present the first systematic empirical study of cross-chain verification vulnerabilities, identifying 11 CVV patterns from 588 audit reports of 36 real-world bridge projects.
    
    \item We construct the first CVV dataset, comprising 830 instances across 786 contract files, and characterize their distribution, severity, and cross-project diversity. The dataset is publicly available at \url{https://anonymous.4open.science/r/CVV-Dataset/}.

    \item We assess the applicability of existing detectors and evaluate four LLMs for CVV detection. Existing tools provide limited coverage, while only two LLMs outperform the 29.3\% Macro-F1 baseline, with the best achieving 51.2\%.
\end{itemize}

%% file: section-02-Background.tex
\section{Background}
\subsection{Cross-chain Bridges}
\begin{figure}[htbp]
    \centering
    \includegraphics[width=0.70\textwidth]{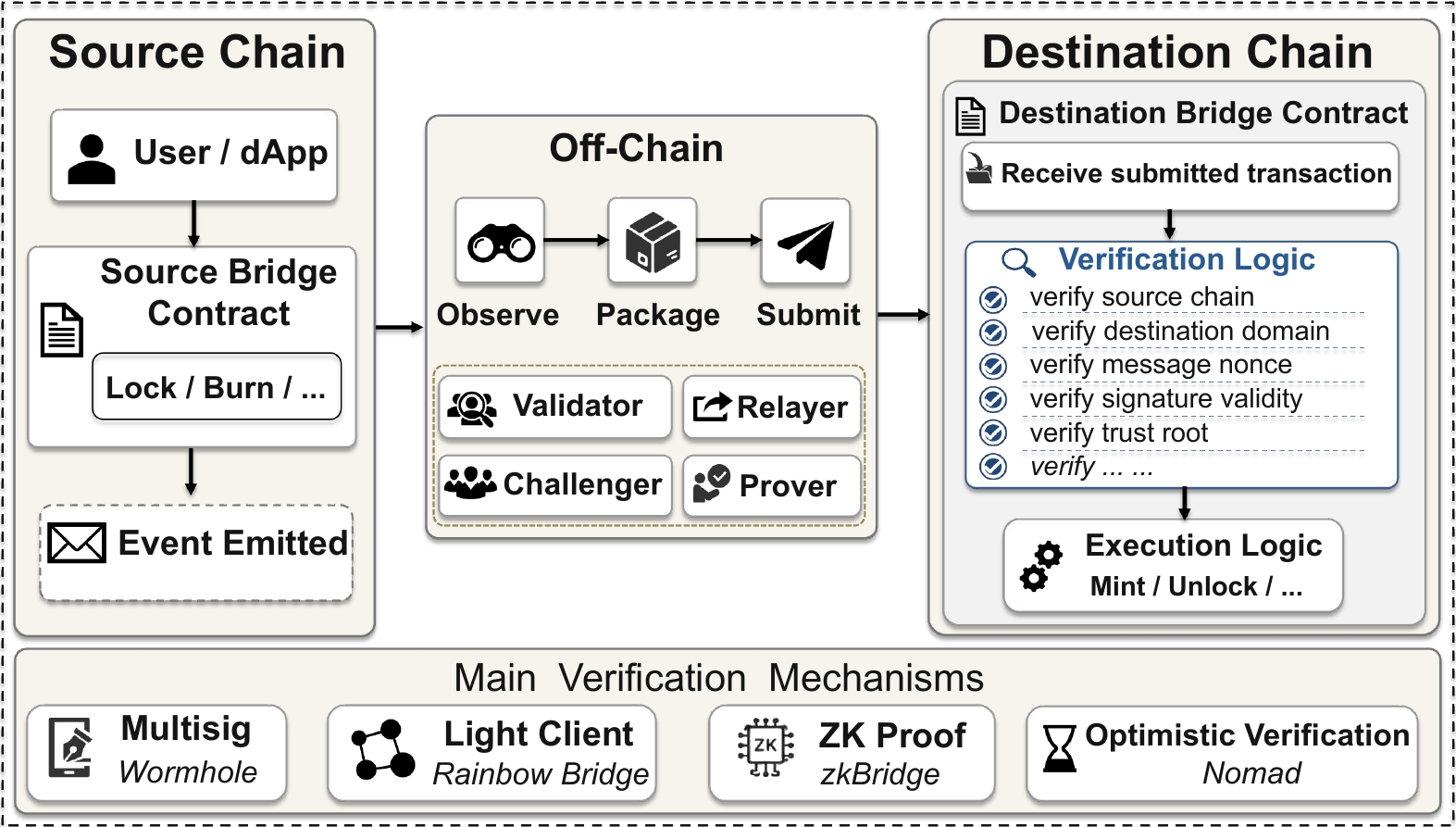}
    \caption{Overview of a Cross-chain Bridge Workflow.}
    \Description{A source-chain user invokes a bridge contract, off-chain actors relay the resulting message or proof, and a destination-chain bridge verifies it before executing an asset transfer or application call.}
    \label{fig:cross-chain-workflow}
    \vspace{-10pt}
\end{figure}

Cross-chain bridges enable interoperability among heterogeneous blockchains through asset transfers, message passing, and state synchronization~\cite{sok-cross-chain}. Fig.~\ref{fig:cross-chain-workflow} shows a typical workflow with a source chain, off-chain actors, and a destination chain. A user or DApp first invokes the source bridge contract, which records the operation as verifiable on-chain data. Off-chain actors collect this data and submit it to the destination chain with the message, signatures, or proof. The destination bridge contract then verifies the data and authorization before executing actions, such as minting assets, releasing funds, or invoking a target contract. Unlike single-chain contracts that rely on local state, bridge contracts must verify external-chain data before executing destination-chain operations. Although some projects, such as Optimism and Polygon, primarily serve Layer-2 scaling, we include their canonical bridge components because they enable interoperability between Ethereum and their corresponding target environments~\cite{ethereum_polygon_bridges}.

\subsection{Cross-chain Verification}
Existing cross-chain bridges use several verification mechanisms. These include multisignature schemes~\cite{multi-signatures}, light clients~\cite{light-clients}, zero-knowledge proofs~\cite{zk}, and optimistic verification~\cite{optimistic} with challenge windows~\cite{verify-mechanism}. Each mechanism defines different evidence that the destination contract must validate. Multisignature bridges, such as Wormhole~\cite{wormhole}, rely on guardian or validator signatures and verify signature thresholds, validator membership, and message-context binding. Light-client bridges, such as Rainbow Bridge~\cite{rainbow}, validate source-chain consensus evidence, including block headers, state roots, Merkle proofs, message inclusion, finality, and freshness. ZK-proof bridges, such as zkBridge~\cite{zkBridge}, verify the public input with a succinct proof to bind a source-chain state or message commitment to a destination-chain operation. Optimistic bridges, such as Nomad~\cite{nomad}, defer acceptance until the challenge window elapses, relying on timeout enforcement, challenge handling, and dispute resolution.

Despite differing trust assumptions and procedures, these mechanisms share security requirements: \textit{cross-chain verification must establish data-origin authenticity and context binding, destination-chain execution integrity, freshness and non-reusability, signature or proof soundness, and trust-root correctness.} This logic spans gateways, messengers, validators, relayers, proof verifiers, decoders, and execution contracts, making it more complex than single-chain execution. Consequently, missing, inconsistent, or incorrectly bound verification logic may cause the destination chain to accept invalid external-chain data and execute unauthorized cross-chain operations.

%% file: section-03-vulnerability-taxonomy.tex
\section{Cross-chain Verification Vulnerabilities}
\label{sec:taxonomy}
\subsection{Data Collection}

To systematically study CVVs, we first collect real-world cross-chain bridge projects and their audit reports. Specifically, our initial data sources consist of bridge projects collected from Chainspot~\cite{chainspot_bridge_db} and DefiLlama~\cite{defillama_bridge_tracker}, two mainstream cross-chain bridge aggregation platforms, with data retrieved as of May 2026. After deduplication and the exclusion of projects without public audit materials, we obtain 36 cross-chain bridge projects.
We then search for publicly available security audit reports for each project. We collect them from project websites, GitHub repositories, and audit-firm websites. The reports involve multiple leading security teams, including Trail of Bits~\cite{trailofbits}, OpenZeppelin~\cite{openzeppelin}, Zellic~\cite{zellic}, Quantstamp~\cite{quantstamp}, and Code4rena~\cite{code4rena}. Because bridge projects are often audited at multiple development stages, we retain all accessible reports to maximize coverage of disclosed vulnerabilities. In total, we collect 588 security audit reports.

Figure~\ref{fig:audit_distribution} shows the distribution of audit reports across different projects. Report counts vary substantially: LayerZero~\cite{layerzero_audits}, Axelar~\cite{axelar_audits}, Optimism~\cite{optimism_audit_reports}, ZKsync~\cite{zksync_audits}, and Arbitrum~\cite{arbitrum_audit_reports} contribute more reports, while several other projects have only a few publicly available audit materials. This distribution reflects differences among cross-chain bridge projects in ecosystem scale, development activity, and transparency of public security audits.

\vspace{-4pt}

\subsection{Data Processing}

\begin{figure}[t]
    \centering

    \begin{subfigure}[t]{0.48\textwidth}
        \centering
        \begin{adjustbox}{
            width=\linewidth,
            height=4.5cm,
            center
        }
            \includegraphics{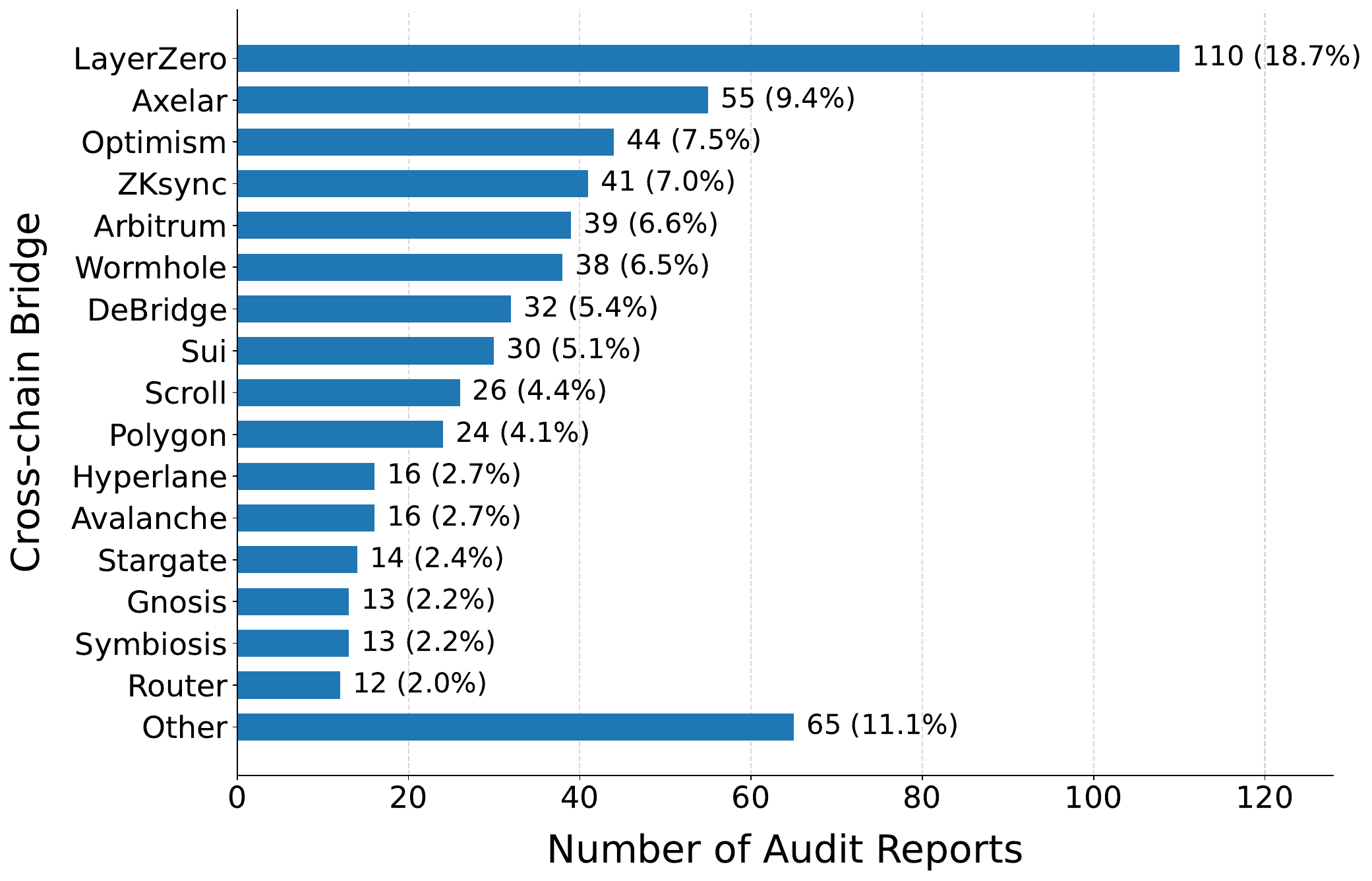}
        \end{adjustbox}
        \caption{Distribution of audit reports by bridge project.}
        \label{fig:audit_distribution}
    \end{subfigure}
    \hfill
    \begin{subfigure}[t]{0.48\textwidth}
        \centering
        \begin{adjustbox}{
            width=\linewidth,
            height=4.3cm,
            center
        }
            \includegraphics{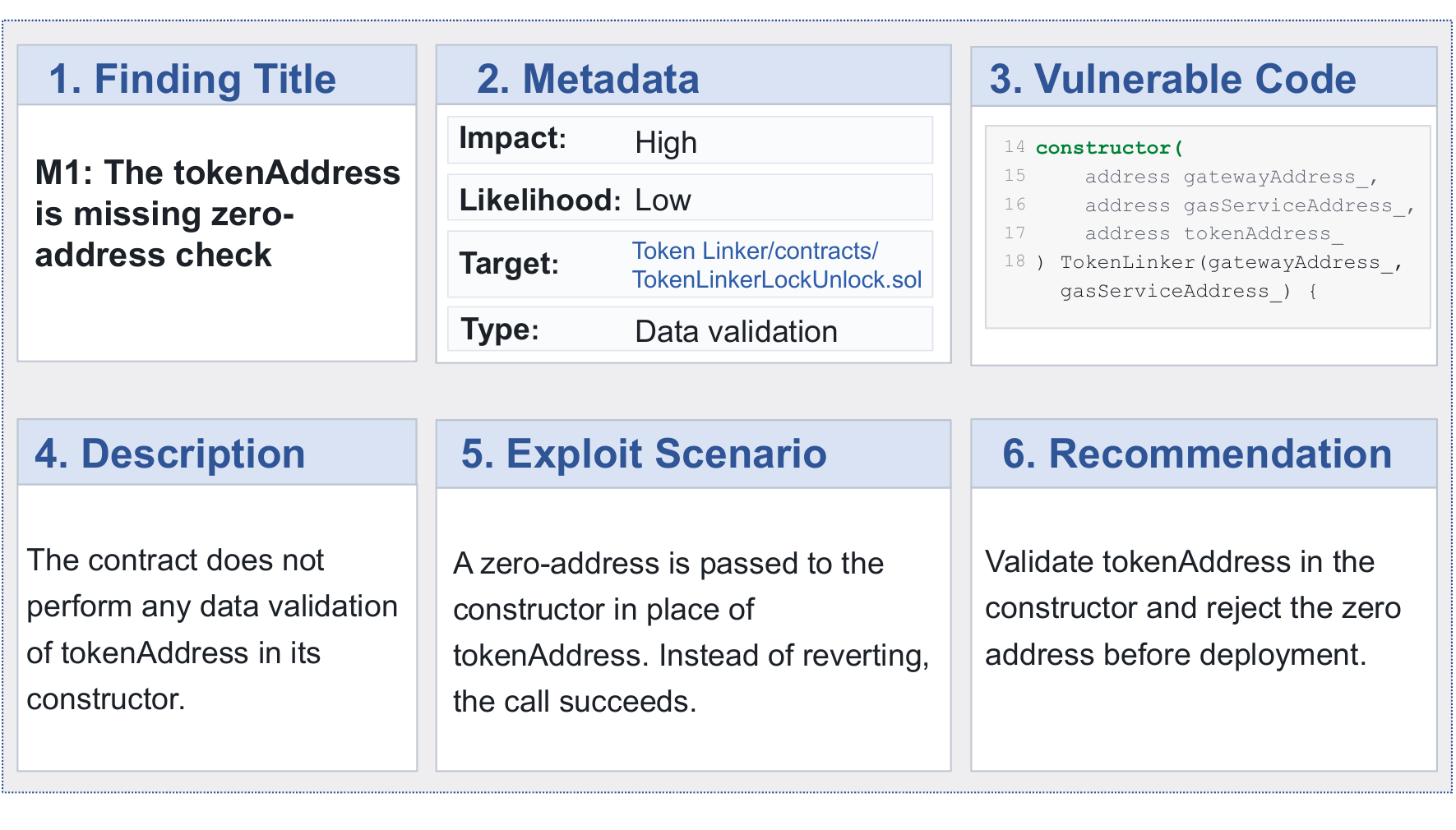}
        \end{adjustbox}
        \caption{Example of a representative audit finding.}
        \label{fig:audit_finding}
    \end{subfigure}

    \caption{Overview of the collected audit reports: project distribution and a representative finding.}
    \Description{The left panel compares the number of collected audit reports across bridge projects; the right panel identifies the title, metadata, vulnerable code, description, exploit scenario, and recommendation in a representative audit finding.}
    \label{fig:audit_overview}
    \vspace{-4pt}
\end{figure}

\begin{table}[t]
\centering
\caption{Audit-finding screening results for CVV identification.}
\label{tab:cvv-screening}
\footnotesize
\renewcommand{\arraystretch}{1.12}
\setlength{\tabcolsep}{4pt}
\begin{tabularx}{0.95\columnwidth}{>{\raggedright\arraybackslash}p{0.36\columnwidth} X r}
\toprule
\textbf{Screening Result} & \textbf{Meaning} & \textbf{Count} \\
\midrule
Fail (i) Cross-chain Relevance
& Unrelated to cross-chain operations
& 3,942 \\

Pass (i), fail (ii) Verification Relevance
& Cross-chain vulnerabilities not involving verification-related objects or consumption of verification results
& 456 \\

Pass (i)--(ii), fail (iii) Verification Impact
& Verification-related findings that do not compromise verification correctness or authorized execution
& 189 \\

Pass (i)--(iii)
& Cross-chain Verification Vulnerabilities (CVVs)
& 1,482 \\
\midrule
\textbf{Total}
& \textbf{Security findings}
& \textbf{6,069} \\
\bottomrule
\end{tabularx}
\vspace{-5pt}
\end{table}

The collected audit reports contain 6,069 security findings spanning diverse implementation defects and security issues. We manually identify CVVs through a three-stage screening procedure using the vulnerable code, descriptions, exploit scenarios, recommendations, and metadata provided in each finding (Figure~\ref{fig:audit_finding}). \textbf{(i) Cross-chain Relevance:} whether the finding affects a cross-chain operation, such as message transmission, asset transfer, state synchronization, verification, or destination-chain execution. \textbf{(ii) Verification Relevance:} whether the affected data, state, or parameters participate in cross-chain verification, including messages, proofs, signatures, state roots, validator sets, identities, and execution parameters. \textbf{(iii) Verification Impact:} whether the defect compromises the correctness of verification or the execution authorized by its result. A finding is labeled as a CVV only if all three criteria are satisfied.

Table~\ref{tab:cvv-screening} summarizes the screening results: among 6,069 findings, 3,942 are excluded as unrelated to cross-chain operations, 456 for lacking verification relevance, and 189 for lacking verification impact, leaving 1,482 CVVs. To reduce labeling bias, two researchers with over three years of smart contract security experience independently assign each finding to one of four screening outcomes, achieving 92.4\% agreement and a Cohen's $\kappa$ of 0.85 before reconciliation~\cite{cohen1960kappa}. Disagreements are resolved by jointly reviewing the audit description, affected code, and security consequences against the screening criteria; findings potentially affected by criterion refinements are re-examined. The reconciled labels form the final CVV set for taxonomy construction and dataset development.

\vspace{-8pt}
\subsection{Data Analysis}


To construct the CVV taxonomy, we apply open card sorting to the 1,482 CVVs. For each instance, we record its description, affected code, triggering condition, exploitation mechanism, security consequence, and remediation. The same two researchers independently group instances by their underlying verification failures, and iteratively merge or split candidate patterns based on verification semantics and cross-chain domain knowledge. Each pattern is characterized by its affected verification object and violated security guarantee, and mapped to one of five guarantees: message authenticity, execution integrity, replay and freshness, verification soundness, or trust management.

After the taxonomy stabilizes, the two researchers independently assign CVV instances to the patterns, achieving 90.8\% agreement and a Cohen's $\kappa$ of 0.88 before reconciliation. Disagreements are resolved through joint review, with ambiguous pattern boundaries refined when necessary. This process results in 5 high-level categories and 11 CVV patterns, as summarized in Table~\ref{tab:cvv-pattern-distribution}.

The 1,482 CVVs span 36 bridge projects and all five verification guarantees. \emph{Improper Handling of Unexpected Input (B3)} is the most frequent pattern, with 705 instances (47.57\%) across 31 projects. B3 captures failures in protocol-level validity checks on cross-chain data, rather than generic local input-validation defects. The remaining patterns cover identity binding, destination validation, replay protection, cryptographic verification, and trust management, showing that CVVs span multiple dimensions of cross-chain verification.


\begin{table}[t]
\centering
\caption{Distribution of CVV instances across 11 fine-grained patterns.}
\label{tab:cvv-pattern-distribution}
\footnotesize
\renewcommand{\arraystretch}{1.12}
\setlength{\tabcolsep}{2.0pt}

\begin{minipage}{0.95\columnwidth}
\centering
\begin{tabular*}{\linewidth}{@{\extracolsep{\fill}}
p{0.22\linewidth}
p{0.37\linewidth}
>{\centering\arraybackslash}p{0.11\linewidth}
>{\centering\arraybackslash}p{0.11\linewidth}
>{\centering\arraybackslash}p{0.08\linewidth}
@{}}
\toprule
\textbf{Category} & \textbf{Pattern} & \textbf{Instances} & \textbf{Percentage} & \textbf{Projects} \\
\midrule
A. Message Authenticity & A1. Missing Message Identity Binding & 115 & 7.76\% & 21 \\
\specialrule{\cmidrulewidth}{\aboverulesep}{\belowrulesep}
\multirow{3}{*}{B. Execution Integrity}
& B1. Missing Execution Parameter Binding & 198 & 13.36\% & 19 \\
& B2. Missing Destination Validation & 108 & 7.29\% & 18 \\
& B3. Improper Handling of Unexpected Input & 705 & 47.57\% & 31 \\
\specialrule{\cmidrulewidth}{\aboverulesep}{\belowrulesep}
\multirow{3}{*}{C. Replay \& Freshness}
& C1. Message/Proof Replay & 109 & 7.35\% & 17 \\
& C2. Signature Replay/Reuse & 35 & 2.36\% & 8 \\
& C3. Stale State Reuse & 23 & 1.55\% & 6 \\
\specialrule{\cmidrulewidth}{\aboverulesep}{\belowrulesep}
\multirow{2}{=}{D. Verification Soundness}
& D1. Improper Signature Verification & 34 & 2.29\% & 11 \\
& D2. Cryptographic Implementation Error & 11 & 0.74\% & 8 \\
\specialrule{\cmidrulewidth}{\aboverulesep}{\belowrulesep}
\multirow{2}{*}{E. Trust Management}
& E1. Improper Trust Root Management & 119 & 8.03\% & 10 \\
& E2. Improper Validator Set Management & 25 & 1.69\% & 12 \\
\midrule
\textbf{Total} & & \textbf{1,482} & \textbf{100.00\%} & \textbf{36} \\
\bottomrule
\end{tabular*}
\end{minipage}
\vspace{-4pt}
\end{table}

\begin{figure}[!t]
    \centering
    \captionsetup{skip=7pt}
    \scalebox{1}[0.82]{\includegraphics[width=\linewidth]{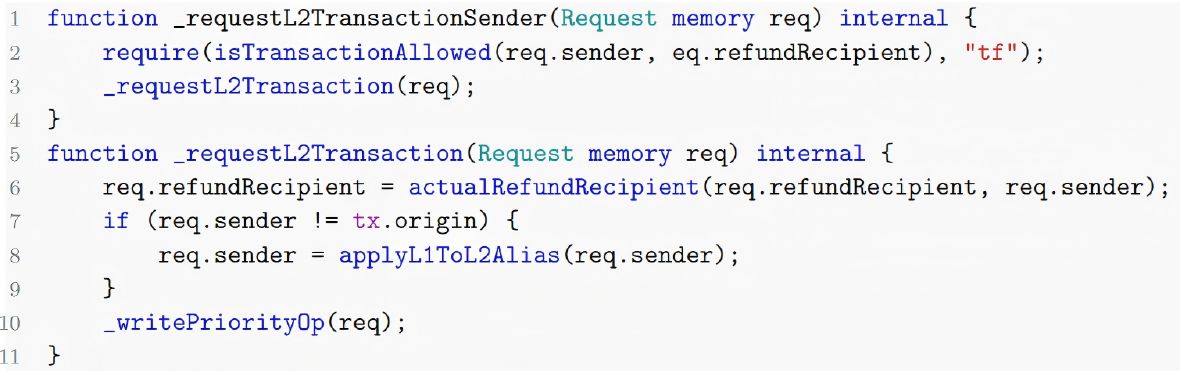}}
    \caption{Example of Missing Message Identity Binding (A1) (source: \texttt{Mailbox.sol:L237--L289}~\cite{a1_code}).}
    \Description{Code excerpt showing a cross-chain mailbox path in which message identity is not fully bound to its source context.}
    \label{fig:pattern-a}
    \par\smallskip
    \scalebox{1}[0.82]{\includegraphics[width=\linewidth]{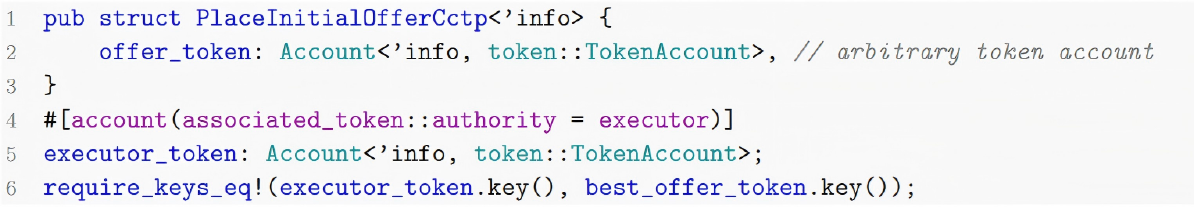}}
    \caption{Example of Missing Execution Parameter Binding (B1) (sources: \texttt{place\_initial/cctp.rs:L13,L95}; \texttt{settle/complete.rs:L17--L22,L115--L117}~\cite{b1_code}).}
    \Description{Two code excerpts illustrating inconsistent binding of a token-account parameter between initial offer placement and later settlement.}
    \label{fig:pattern-b}
    \vspace{-5pt}
\end{figure}

\subsection{Vulnerability Taxonomy} 
We classify CVVs according to the cross-chain verification security guarantees they violate. Specifically, in Table~\ref{tab:cvv-pattern-distribution}, we organize CVVs into 5 high-level categories: \emph{Message Authenticity Violation}, \emph{Execution Integrity Violation}, \emph{Replay and Freshness Violation}, \emph{Verification Soundness Violation}, and \emph{Trust Management Violation}. Under these categories, we further define 11 concrete CVV patterns. 

\noindent\textbf{(1) Message Authenticity Violation (CVV-A).}
Message authenticity requires the destination-chain contract to reliably identify the origin of a cross-chain message, including its source chain, source contract, sender, domain, and unique message identifier. This category contains one pattern: \emph{Missing Message Identity Binding (A1)}, where a message is not fully bound to its source context. As a result, the destination-chain contract may fail to distinguish the legitimate origin of a message from a forged or mismatched context.

\textbf{Example:} As shown in Fig.~\ref{fig:pattern-a}, the destination-side contract verifies whether the source-side \code{sender} and \code{refundRecipient} are allowed. Before destination-chain execution, both identities may be transformed through address aliasing. This mismatch between the verified source-side identity and the executed destination-side identity violates message identity binding and may allow restricted cross-chain requests to bypass access control.

\noindent\textbf{(2) Execution Integrity Violation (CVV-B).} 
Execution integrity requires cross-chain data to satisfy the protocol-level constraints necessary for safe execution and, once verified, to be executed consistently with the verified context, including its parameters, destination, and intended behavior. This category includes three patterns. 
\emph{Missing Execution Parameter Binding (B1)} occurs when execution-critical parameters are not bound to the verification context. 
\emph{Missing Destination Validation (B2)} occurs when the destination chain, target contract, or target function is not properly validated. 
\emph{Improper Handling of Unexpected Input (B3)} occurs when malformed, unsupported, or otherwise invalid cross-chain inputs are not rejected by the required validity checks and are subsequently accepted or consumed for execution, causing destination-chain behavior to deviate from the intended cross-chain semantics.

\textbf{Example:} As shown in Fig.~\ref{fig:pattern-b}, \code{PlaceInitialOfferCctp()} accepts \code{offer_token} as an arbitrary token account. However, the execution constraint later requires \code{executor_token.key()} to match \code{best_offer_token.key()}. This inconsistent token-account binding shows that execution-critical parameters must be consistently constrained across stages.

\begin{figure}[!t]
    \centering
    \captionsetup{skip=7pt}
    \scalebox{1}[0.82]{\includegraphics[width=\linewidth]{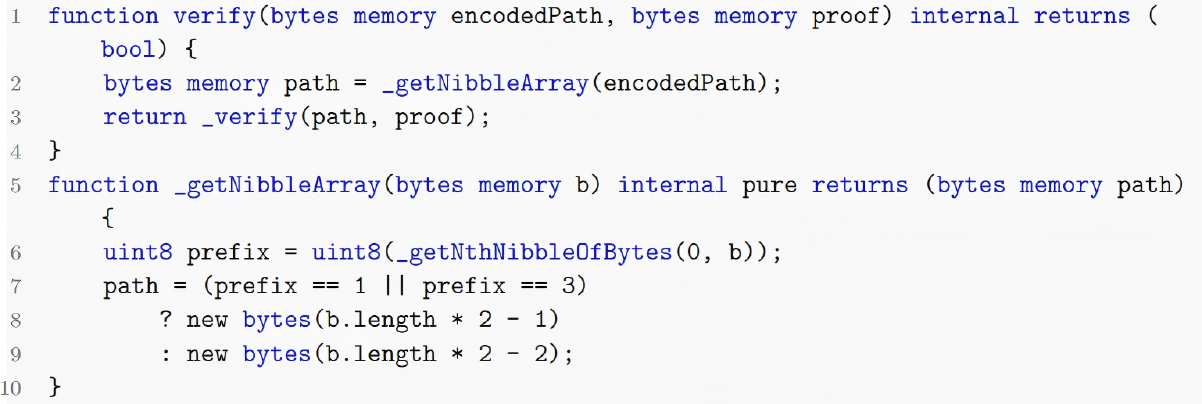}}
    \caption{Example of Message/Proof Replay (C1) (source: \texttt{MerklePatriciaProof.sol:L20--36,136--152}~\cite{c1_code}).}
    \Description{Code excerpt showing how distinct encodings can resolve to the same Merkle proof path while producing different request identifiers, enabling replay.}
    \label{fig:pattern-c}
    \par\smallskip
    \scalebox{1}[0.82]{\includegraphics[width=\linewidth]{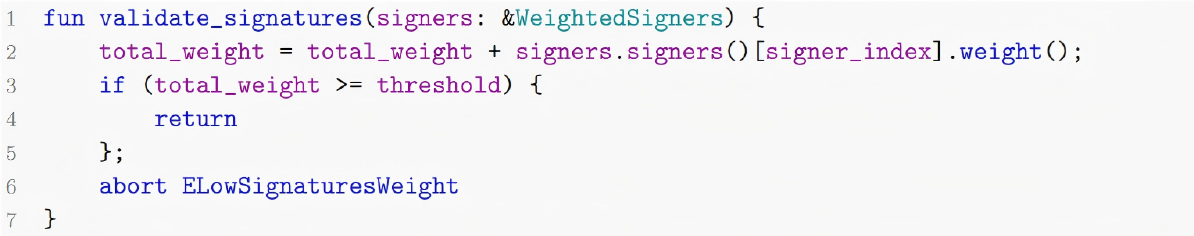}}
    \caption{Example of Improper Signature Verification (D1) (source: \texttt{validators.move:L74--L95}~\cite{d1_code}).}
    \Description{Code excerpt illustrating signature-verification logic whose validity depends on an improperly enforced signature ordering rule.}
    \label{fig:pattern-d}
    \par\smallskip
    \scalebox{1}[0.82]{\includegraphics[width=\linewidth]{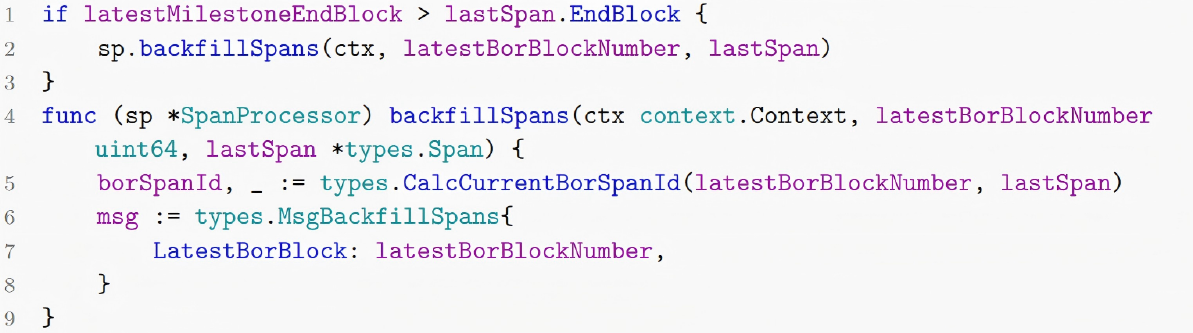}}
    \caption{Example of Improper Trust Root Management (E1) (source: \texttt{span.go:L114--L116,L139--L188}~\cite{e1_code}).}
    \Description{Code excerpt showing a backfill operation based on a latest block number that may not yet be finalized, allowing rollback-prone state to become a verification basis.}
    \label{fig:pattern-e}
\end{figure}

\noindent\textbf{(3) Replay and Freshness Violation (CVV-C).} 
Replay and freshness require cross-chain messages, proofs, signatures, and states to be unique, fresh, and single-use after verification. This category includes three patterns. \emph{Message/Proof Replay (C1)} occurs when a message or proof lacks uniqueness or reuse-prevention verification. \emph{Signature Replay/Reuse (C2)} occurs when a signature is not bound to a unique context, such as chain ID, contract address, nonce, or expiration. \emph{Stale State Reuse (C3)} occurs when outdated states, such as stale state roots or expired validator sets, are still accepted.

\textbf{Example:} As shown in Fig.~\ref{fig:pattern-c}, the same burn proof can be submitted with different encodedPath encodings. Since \code{_getNibbleArray()} derives the verification path from the first nibble without strictly validating the path prefix, different encodings may be parsed into the same path and pass verification against the same proof. However, they can still produce different request identifiers, causing the contract to treat the same proof as distinct requests and enabling proof replay.

\noindent\textbf{(4) Verification Soundness Violation (CVV-D).} Verification soundness requires cryptographic and logical checks in cross-chain verification to be correctly implemented. This category includes two patterns. \emph{Improper Signature Verification (D1)} occurs when the logic fails to bind the message, signature, signer, and verification context. \emph{Cryptographic Implementation Error (D2)} occurs when cryptographic mechanisms are incorrectly implemented, such as using wrong hash encodings, invalid proofs, or incorrect signature recovery parameters.

\textbf{Example:} As shown in Fig.~\ref{fig:pattern-d}, \code{validate_signatures()} accepts a message once the accumulated signer weight reaches \code{threshold}. If \code{threshold} is zero, the condition \code{total_weight >= threshold} can be satisfied trivially, allowing signatures with insufficient or even zero validator weight to pass verification. This bypasses the intended validator-weight consensus and violates the soundness of cross-chain signature verification.

\noindent\textbf{(5) Trust Management Violation (CVV-E).} Trust management requires the trust roots, validator sets, and related states used for verification to be valid, finalized, and current. This category includes two patterns. \emph{Improper Trust Root Management (E1)} occurs when stale, non-finalized, unauthorized, or untrusted state roots are used for verification. \emph{Improper Validator Set Management (E2)} allows invalid, expired, or insufficiently weighted validators to participate in verification.

\textbf{Example:} As shown in Fig.~\ref{fig:pattern-e}, the contract decides whether to backfill based on \code{latestMilestoneEndBlock}, representing finalized progress. However, it passes \code{latestBorBlockNumber} to \code{backfillSpans()}, although this block may not be finalized. Consequently, the system may derive verification states from rollback-prone blocks and treat non-finalized states as verification bases.

\begin{figure*}[!t]
    \centering
    \scalebox{1}[0.97]{\includegraphics[width=0.93\textwidth]{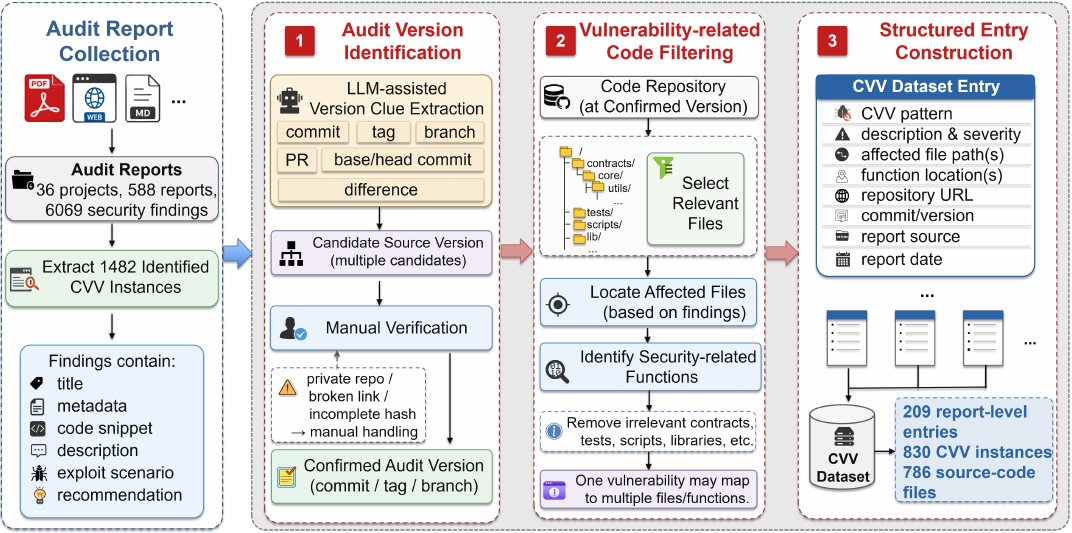}}
    \caption{Workflow for Constructing the Cross-chain Verification Vulnerability Dataset.}
    \Description{A three-stage workflow that identifies audited source-code versions, filters vulnerability-related code, and constructs structured dataset entries linking reports, versions, files, and CVV labels.}
    \label{fig:dataset-construction}
    \vspace{-5pt}
\end{figure*}

%% file: section-04-vulnerability-dataset.tex
\section{Dataset Construction and Characterization}
\subsection{Dataset Construction}

Based on the taxonomy in Section~\ref{sec:taxonomy}, we map each CVV instance to its audited source-code version and affected files. Figure~\ref{fig:dataset-construction} summarizes the three-step workflow: audited-version identification, vulnerability-related code filtering, and structured-entry construction.

For each CVV instance, we first recover the audited source-code version. Because audit reports may reference commits, tags, branches, pull requests, or commit ranges, we use GPT-5.5 only to extract version clues, which are then manually verified against report descriptions, affected files, code snippets, and patch records. Instances whose audited versions cannot be reliably recovered are excluded. This process retains 830 of the 1,482 CVVs from 209 audit reports across 27 bridge projects; the extraction prompt is included in our replication package.

We then locate the vulnerability-related code in each confirmed version, retaining all affected source files while excluding irrelevant tests, scripts, libraries, and contracts. The dataset is organized at the audit-report level, with each entry recording repository and version information, report metadata, and all associated CVVs, including their patterns, descriptions, severities, and affected files. These mappings support subsequent characterization and detection experiments.
\begin{table}[!t]
\centering
\footnotesize
\caption{Statistics of the CVV dataset.}
\label{tab:dataset_statistics}
\renewcommand{\arraystretch}{1.08}
\setlength{\tabcolsep}{4pt}

\begin{tabularx}{0.82\columnwidth}{Xr}
\toprule
\textbf{Statistic} & \textbf{Value} \\
\midrule
Security audit reports used for dataset construction & 209 \\
Cross-chain bridge projects covered by the dataset & 27 \\
Cross-chain verification vulnerability instances & 830 \\
Source-code files mapped to CVV instances & 786 \\
\addlinespace[2pt]
CVV instances per dataset entry & 3.97; 2 [1--5] \\
Source-code files per dataset entry & 3.76; 3 [2--5] \\
\midrule
Solidity source-code files & 490 \\
Rust source-code files & 159 \\
Go source-code files & 76 \\
Source-code files written in other languages & 61 \\
\bottomrule
\end{tabularx}

\vspace{2pt}
\begin{minipage}{0.82\columnwidth}
\scriptsize
\textit{Note:} Per-entry statistics are reported as mean; median [Q1--Q3].
\end{minipage}
\vspace{-4pt}
\end{table}

\subsection{Dataset Overview}

Table~\ref{tab:dataset_statistics} summarizes our CVV dataset. Of the 1,482 CVVs identified in Section~\ref{sec:taxonomy}, 830 instances with reliably recovered audited versions are retained, covering 209 audit reports and 27 bridge projects. Each dataset entry corresponds to one report and includes all retained CVVs, which are mapped to 786 source-code files. Each report contains an average of 3.97 CVVs (median: 2, IQR: 1--5) and 3.76 files (median: 3, IQR: 2--5), indicating right-skewed distributions. Using the 1.5$\times$IQR criterion, we identify 13 and 10 high-end outliers, respectively, all retained as valid audit cases. The dataset includes 490 Solidity, 159 Rust, 76 Go, and 61 other-language files, spanning heterogeneous bridge implementations and report scales.
\subsection{Vulnerability Characterization}
\subsubsection{Vulnerability Pattern Distribution}

Table~\ref{tab:cvv_pattern_distribution} summarizes the distribution of 830 CVVs across 5 categories and 11 patterns. The distribution is highly imbalanced: \emph{Execution Integrity} accounts for 577 instances (69.52\%), followed by \emph{Trust Management} with 87 (10.48\%), while \emph{Verification Soundness} is the least frequent with 21 (2.53\%).
At the pattern level, \emph{Improper Handling of Unexpected Input (B3)} is the most prevalent, with 409 instances (49.28\%), followed by \emph{Missing Execution Parameter Binding (B1)} with 100 (12.05\%). 
This concentration shows that many real-world CVVs arise in how authenticated cross-chain data is interpreted, constrained, and consumed during destination-chain execution, rather than in cryptographic verification alone.

\finding{1}{69.52\% of CVVs violate execution integrity, indicating that effective detection must model the verification-to-execution path rather than focus only on cryptographic checks.}

\begin{table}[!t]
\centering
\footnotesize
\caption{Distribution of CVV categories and patterns.}
\label{tab:cvv_pattern_distribution}
\renewcommand{\arraystretch}{1.12}
\setlength{\tabcolsep}{2pt}
\begin{minipage}{0.95\columnwidth}
\centering
\begin{tabular*}{\linewidth}{@{\extracolsep{\fill}}
>{\raggedright\arraybackslash}p{0.22\linewidth}
>{\centering\arraybackslash}p{0.15\linewidth}
>{\raggedright\arraybackslash}p{0.40\linewidth}
>{\centering\arraybackslash}p{0.14\linewidth}
@{}}
\toprule
\textbf{Category} & \textbf{Instances (\%)} & \textbf{Pattern} & \textbf{Instances (\%)} \\
\midrule
A. Message Authenticity
& 63 (7.59)
& A1. Missing Message Identity Binding
& 63 (7.59) \\
\specialrule{\cmidrulewidth}{\aboverulesep}{\belowrulesep}
\multirow{3}{=}{B. Execution Integrity}
& \multirow{3}{=}{\centering 577 (69.52)}
& B1. Missing Execution Parameter Binding & 100 (12.05) \\
& & B2. Missing Destination Validation & 68 (8.19) \\
& & B3. Improper Handling of Unexpected Input & 409 (49.28) \\
\specialrule{\cmidrulewidth}{\aboverulesep}{\belowrulesep}
\multirow{3}{=}{C. Replay \& Freshness}
& \multirow{3}{=}{\centering 82 (9.88)}
& C1. Message/Proof Replay & 57 (6.87) \\
& & C2. Signature Replay/Reuse & 10 (1.20) \\
& & C3. Stale State Reuse & 15 (1.81) \\
\specialrule{\cmidrulewidth}{\aboverulesep}{\belowrulesep}
\multirow{2}{=}{D. Verification Soundness}
& \multirow{2}{=}{\centering 21 (2.53)}
& D1. Improper Signature Verification & 15 (1.81) \\
& & D2. Cryptographic Implementation Error & 6 (0.72) \\
\specialrule{\cmidrulewidth}{\aboverulesep}{\belowrulesep}
\multirow{2}{=}{E. Trust Management}
& \multirow{2}{=}{\centering 87 (10.48)}
& E1. Improper Trust Root Management & 75 (9.04) \\
& & E2. Improper Validator Set Management & 12 (1.45) \\
\midrule
\textbf{Total} & \textbf{830 (100.00)} & & \textbf{830 (100.00)} \\
\bottomrule
\end{tabular*}
\end{minipage}
\vspace{-4pt}
\end{table}

\subsubsection{Vulnerability Severity Analysis}
Since audit firms use heterogeneous severity labels, we normalize them into four levels---\emph{Low}, \emph{Medium}, \emph{High}, and \emph{Informational}---for cross-report comparison (Table~\ref{tab:severity_distribution}). Semantically equivalent labels are mapped to the same level, while labels without explicit severity indications are treated as \emph{Unspecified}. Overall, 199 CVVs (23.98\%) are labeled \emph{High}.


\begin{table}[!t]
\centering
\footnotesize
\caption{Severity normalization and distribution of CVV instances.}
\label{tab:severity_distribution}
\renewcommand{\arraystretch}{1.12}
\setlength{\tabcolsep}{2pt}

\begin{minipage}{0.95\columnwidth}
\centering
\begin{tabularx}{\linewidth}{
@{}
>{\raggedright\arraybackslash}p{0.18\linewidth}
>{\raggedright\arraybackslash}X
>{\centering\arraybackslash}p{0.12\linewidth}
>{\centering\arraybackslash}p{0.14\linewidth}
@{}}
\toprule
\textbf{Severity} &
\textbf{Original Labels} &
\textbf{Instances} &
\textbf{Percentage} \\
\midrule

Low
& Low, Minor, Low Risk
& 251 & 30.24\% \\

Medium
& Medium, Major, Moderate
& 214 & 25.78\% \\

High
& Critical, High, Severe, High Risk
& 199 & 23.98\% \\

Informational
& Informational, Info, Warning, Suggestion
& 147 & 17.71\% \\

Unspecified
& Status, Acknowledged, other non-severity labels
& 19 & 2.29\% \\

\midrule
\textbf{Total}
& &
\textbf{830}
& \textbf{100.00\%} \\

\bottomrule
\end{tabularx}
\end{minipage}
\vspace{-6pt}
\end{table}

\begin{figure}[tbp]
    \centering
    \scalebox{1.08}[0.96]{\includegraphics[width=0.62\textwidth]{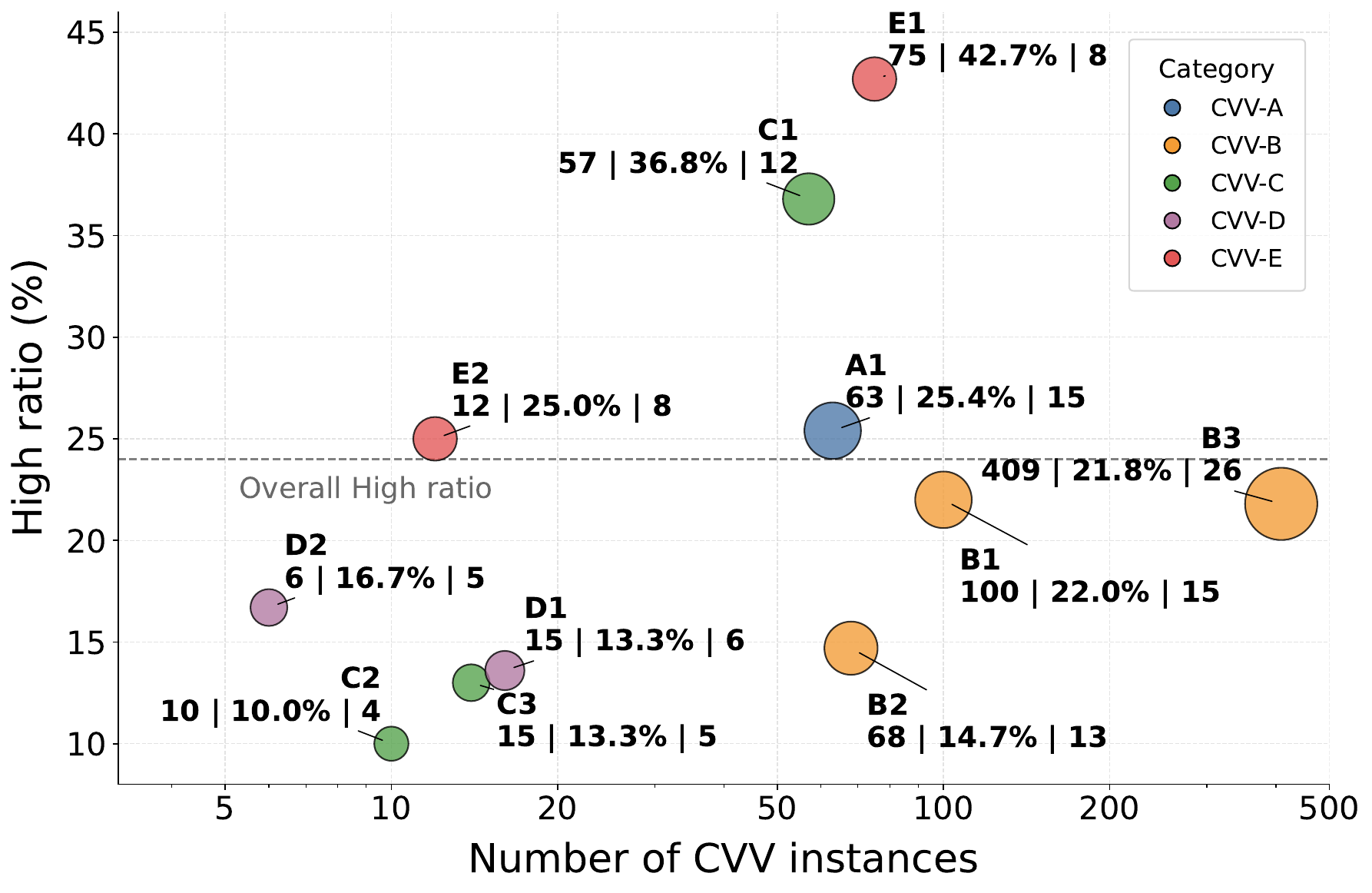}}
    \caption{Risk distribution of CVV patterns. Bubble labels follow the format $N \mid H\% \mid B$, where $N$, $H\%$, and $B$ denote the number of instances, the High ratio, and the number of affected bridges, respectively.}
    \Description{Bubble chart comparing CVV patterns by instance count, percentage of High-severity findings, and number of affected bridge projects.}
    \label{fig:vul-Severity}
    \vspace{-6pt}
\end{figure}

Figure~\ref{fig:vul-Severity} compares CVV patterns by prevalence, severity, and affected-project scope. \emph{Improper Handling of Unexpected Input (B3)} is the most prevalent and widespread pattern, but not the most severe. In contrast, \emph{Improper Trust Root Management (E1)} and \emph{Message/Proof Replay (C1)} occur less frequently, with 75 and 57 instances, but reach \emph{High}-severity ratios of 42.7\% and 36.8\%, respectively. These results show that prevalence does not necessarily correspond to severity.

\finding{2}{CVVs pose substantial security risks: 23.98\% are labeled High, while less prevalent E1 and C1 reach 42.7\% and 36.8\%, showing that lower frequency does not imply lower impact.}


\subsubsection{Vulnerability Distribution Across Cross-chain Bridges}

To examine how CVVs are distributed across bridge projects, we analyze the six most represented projects in the retained dataset in terms of instance counts and pattern composition. As shown in Table~\ref{tab:bridge_distribution}, each project involves at least six CVV patterns: Polygon and Wormhole cover all 11 patterns, while Optimism, Axelar, Scroll, and ZKsync cover 6--9. This indicates that CVVs span multiple verification dimensions across heterogeneous bridge implementations rather than being confined to a single project or pattern.

\begin{figure}[!t]
    \centering
    \scalebox{1.08}[0.96]{\includegraphics[width=0.62\textwidth]{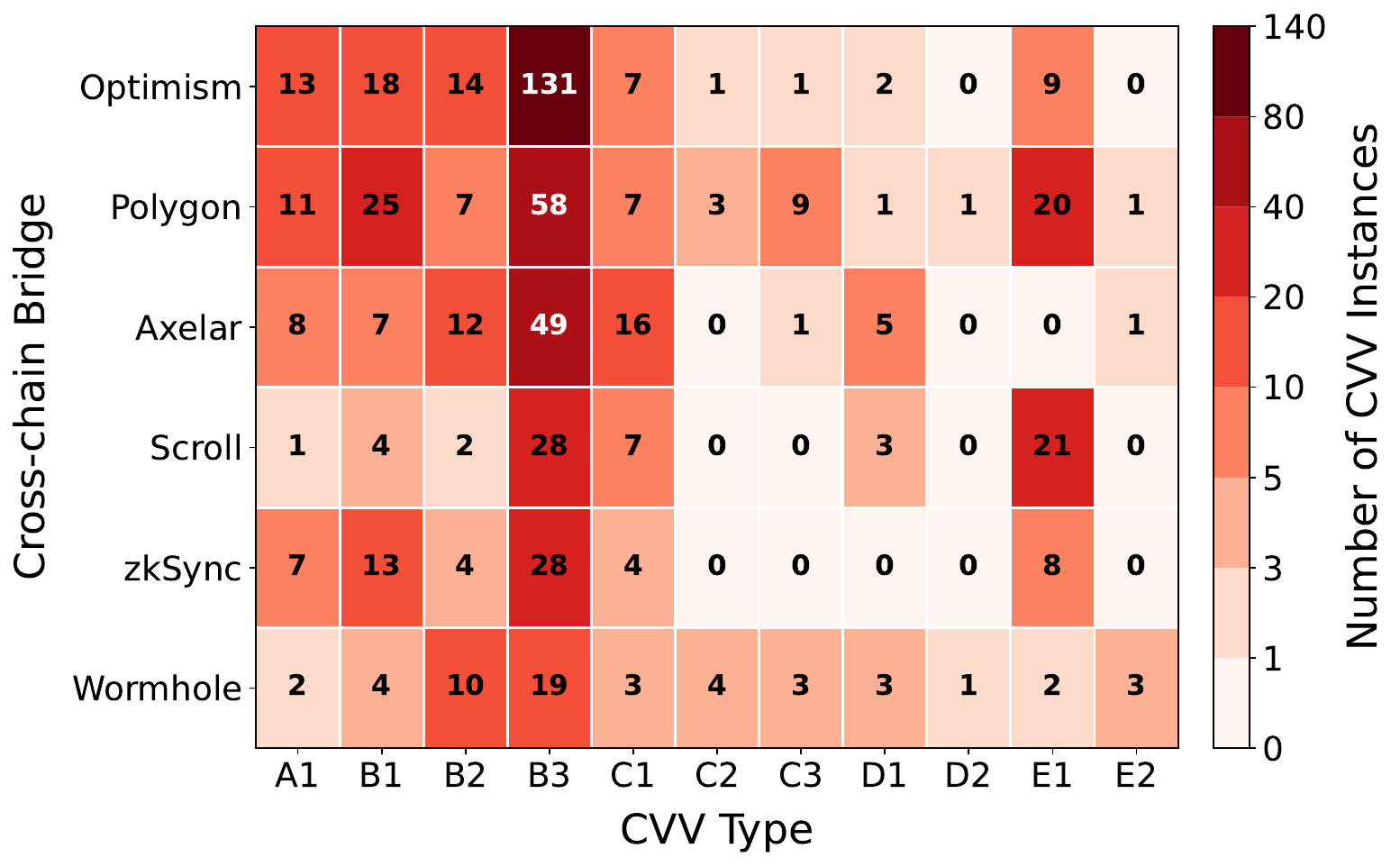}}
    \caption{CVV Pattern Distribution Across the Top-6 Cross-chain Bridges.}
    \Description{Heat map comparing the counts of eleven CVV patterns across Polygon, Optimism, Axelar, Scroll, ZKsync, and Wormhole.}
    \label{fig:cross-chain bridges}
\end{figure}

Figure~\ref{fig:cross-chain bridges} further shows the CVV pattern composition of these projects. \emph{Improper Handling of Unexpected Input (B3)} is prevalent across all six, accounting for 313 of 622 CVVs (50.32\%). Meanwhile, the remaining patterns exhibit distinct project-specific compositions: Polygon shows broader exposure to B1, E1, and A1; Axelar to C1 and B2; Scroll to E1; and Wormhole has a more dispersed pattern distribution.

\begin{table}[!t]
\centering
\footnotesize
\caption{Top six cross-chain bridges by number of CVV instances.}
\label{tab:bridge_distribution}
\renewcommand{\arraystretch}{1.12}
\setlength{\tabcolsep}{9pt}
\begin{tabular}{@{}cccc@{}}
\toprule
\textbf{Cross-chain Bridge} & \textbf{Instances} & \textbf{Percentage} & \textbf{CVV Patterns} \\
\midrule
Optimism & 196 & 23.61\% & 9 \\
Polygon & 143 & 17.23\% & 11 \\
Axelar & 99 & 11.93\% & 8 \\
Scroll & 66 & 7.95\% & 7 \\
ZKsync & 64 & 7.71\% & 6 \\
Wormhole & 54 & 6.51\% & 11 \\
\bottomrule
\end{tabular}
\vspace{-6pt}
\end{table}

\finding{3}{Each of the six most represented projects covers at least six CVV patterns, while B3 accounts for 50.32\% of their CVVs, revealing both broad cross-project prevalence and distinct project-specific compositions.}

%% file: section-05-vulnerability-detection-evaluation.tex
\section{Vulnerability Detection Evaluation}
In this section, we evaluate the detectability of CVVs and answer the following research questions:

\begin{itemize}
\item \textbf{RQ1. How applicable are existing smart contract vulnerability detectors to CVV detection?}
We assess their inputs, cross-contract analysis capabilities, and coverage of CVV patterns.
\item \textbf{RQ2. How effectively can LLMs detect CVVs?}
We evaluate mainstream LLMs using precision, recall, and F1 score, and compare their performance across CVV patterns.
\item \textbf{RQ3. What are the main causes of LLM-based detection errors?} 
We manually analyze false negatives and false positives to identify the main limitations of current LLM-based CVV detection.

\end{itemize}

\subsection{RQ1: Applicability of Existing Smart Contract Vulnerability Detection Tools}

To answer RQ1, we assess the documented applicability of 12 representative tools to CVV detection, including 8 traditional single-chain tools mainly from SmartBugs~\cite{smartbugs} and 4 cross-chain-related tools. As summarized in Table~\ref{tab:non_llm_tool_applicability}, these tools span static analysis, rule matching, symbolic execution, fuzzing, and bridge-specific analysis. We evaluate them along three dimensions: supported input, analysis scope, and CVV pattern coverage.

\begin{table}[!t]
\centering
\footnotesize
\caption{Applicability of existing vulnerability detection tools to CVV detection.}
\label{tab:non_llm_tool_applicability}
\renewcommand{\arraystretch}{1.06}
\setlength{\tabcolsep}{2.5pt}
\begin{minipage}{0.95\columnwidth}
\centering
\begin{tabularx}{\linewidth}{
@{\hspace{4pt}}
>{\raggedright\arraybackslash}p{0.17\linewidth}
>{\raggedright\arraybackslash}p{0.16\linewidth}
>{\raggedright\arraybackslash}p{0.22\linewidth}
>{\raggedright\arraybackslash}X
>{\centering\arraybackslash}p{0.16\linewidth}
@{}}
\toprule
\textbf{Tool} & \textbf{Input} & \textbf{Technique} & \textbf{Analysis Scope} & \textbf{Mapped Patterns} \\
\midrule
Slither~\cite{Slither} & Solidity & Static analysis & Single contract & 0/11 \\
SmartCheck~\cite{SmartCheck} & Solidity & Rule matching & Single contract & 0/11 \\
Solhint~\cite{solhint} & Solidity & Linting & Single contract & 0/11 \\
Mythril~\cite{mythril} & EVM bytecode & Symbolic execution & Limited multi-contract & 0/11 \\
Securify~\cite{Securify} & EVM bytecode & Pattern analysis & Single contract & 0/11 \\
ConFuzzius~\cite{ConFuzzius} & Solidity & Hybrid fuzzing & Single contract & 0/11 \\
Oyente~\cite{Oyente} & EVM bytecode & Symbolic execution & Single contract & 0/11 \\
Manticore~\cite{Manticore} & EVM bytecode & Symbolic execution & Limited multi-contract & 0/11 \\
\specialrule{\cmidrulewidth}{\aboverulesep}{\belowrulesep}
SmartAxe~\cite{smartaxe} & EVM bytecode & Static analysis & Cross-contract & 2/11 \\
BridgeGuard~\cite{bridgeguard} & EVM bytecode & Symbolic dataflow & Router interactions & 1/11 \\
ChainSniper~\cite{ChainSniper} & Source code & Machine learning & Contract classification & 0/11 \\
XGuard~\cite{XGuard} & Solidity & Behavior consistency & Cross-chain behavior & 1/11 \\
\bottomrule
\end{tabularx}
\end{minipage}
\end{table}

\textbf{Tool input.}
Most tools analyze Solidity source files or EVM bytecode, whereas real-world bridge systems may additionally involve non-EVM components written in Rust or Go, as well as proof generators and validator services. Bytecode-based tools also lose part of the source-level semantics needed to interpret verification-specific logic. These heterogeneous input requirements limit the applicability of existing tools to CVV analysis.

\textbf{Analysis scope.}
CVVs often involve interactions across multiple contracts or bridge components, requiring reasoning beyond independent contract analysis. As shown in Table~\ref{tab:non_llm_tool_applicability}, most traditional tools focus on individual contracts. Mythril and Manticore provide limited multi-contract execution modeling but primarily target generic EVM vulnerabilities. Among cross-chain-related tools, SmartAxe and XGuard support cross-contract or cross-chain semantic analysis, whereas BridgeGuard and ChainSniper focus on more specific interaction or classification scenarios.

\textbf{CVV pattern coverage.}
We manually map each tool's documented vulnerability scope and analysis semantics to the 11 CVV patterns based on its published paper and documentation. A pattern is mapped only when the tool explicitly targets the corresponding vulnerability or directly captures its underlying verification failure. Under this criterion, none of the traditional tools maps directly to any CVV pattern; among cross-chain tools, SmartAxe maps to 2 patterns, BridgeGuard and XGuard to 1 each, and ChainSniper to none.
\finding{4}{Existing tools have limited applicability to CVV detection, constrained by heterogeneous inputs, cross-component reasoning, and sparse pattern coverage.}

\begin{table}[t]
\centering
\caption{Detection results of four LLMs across 11 CVV patterns (\%).}
\label{tab:rq2_llm_results}
\footnotesize
\setlength{\tabcolsep}{1.5pt}
\renewcommand{\arraystretch}{0.90}
\begin{tabular*}{\columnwidth}{@{\hspace{4pt}\extracolsep{\fill}}c*{12}{c}@{}}
\toprule
\multirow{2}{*}{\textbf{Pattern}}
& \multicolumn{3}{c}{\textbf{GPT-5.5}}
& \multicolumn{3}{c}{\textbf{Composer-2.5}}
& \multicolumn{3}{c}{\textbf{Opus-4.8}}
& \multicolumn{3}{c}{\textbf{DeepSeek-V4-Pro}} \\
\cmidrule(lr){2-4}\cmidrule(lr){5-7}\cmidrule(lr){8-10}\cmidrule(l){11-13}
& \textbf{Prec.} & \textbf{Rec.} & \textbf{F1}
& \textbf{Prec.} & \textbf{Rec.} & \textbf{F1}
& \textbf{Prec.} & \textbf{Rec.} & \textbf{F1}
& \textbf{Prec.} & \textbf{Rec.} & \textbf{F1} \\
\midrule
A1 & 53.8 & 29.2 & 37.8 & 38.3 & 64.6 & 48.1 & 28.6 & 16.7 & 21.1 & 29.9 & 41.7 & 34.8 \\
B1 & 57.1 & 36.9 & 44.9 & 42.0 & 92.3 & 57.7 & 38.5 & 7.7  & 12.8 & 42.6 & 40.0 & 41.3 \\
B2 & 68.2 & 28.3 & 40.0 & 56.9 & 77.4 & 65.6 & 56.2 & 17.0 & 26.1 & 31.9 & 41.5 & 36.1 \\
B3 & 92.8 & 42.7 & 58.4 & 85.3 & 92.7 & 88.8 & 87.8 & 24.0 & 37.7 & 79.7 & 31.3 & 45.0 \\
C1 & 76.2 & 35.6 & 48.5 & 50.8 & 66.7 & 57.7 & 25.0 & 6.7  & 10.5 & 27.0 & 44.4 & 33.6 \\
C2 & 15.4 & 25.0 & 19.0 & 25.0 & 50.0 & 33.3 & 25.0 & 25.0 & 25.0 & 0.0  & 0.0  & 0.0 \\
C3 & 16.7 & 30.0 & 21.4 & 14.5 & 90.0 & 25.0 & 30.0 & 30.0 & 30.0 & 12.5 & 60.0 & 20.7 \\
D1 & 50.0 & 23.1 & 31.6 & 40.9 & 69.2 & 51.4 & 20.0 & 7.7  & 11.1 & 0.0  & 0.0  & 0.0 \\
D2 & 0.0  & 0.0  & 0.0  & 14.3 & 50.0 & 22.2 & 22.2 & 33.3 & 26.7 & 14.3 & 16.7 & 15.4 \\
E1 & 37.9 & 26.8 & 31.4 & 37.9 & 95.1 & 54.2 & 58.3 & 17.1 & 26.4 & 23.3 & 17.1 & 19.7 \\
E2 & 31.6 & 50.0 & 38.7 & 45.5 & 83.3 & 58.8 & 0.0  & 0.0  & 0.0  & 21.9 & 58.3 & 31.8 \\
\midrule
\textbf{Macro Avg.}
& \textbf{45.4} & \textbf{29.8} & \textbf{33.8}
& \textbf{41.0} & \textbf{75.6} & \textbf{51.2}
& \textbf{35.6} & \textbf{16.8} & \textbf{20.7}
& \textbf{25.7} & \textbf{31.9} & \textbf{25.3} \\
\bottomrule
\end{tabular*}
\end{table}

\begin{table}[t]
\centering
\caption{Constant-baseline comparisons at model and pattern levels. $\Delta$ denotes the F1 difference from the corresponding baseline in percentage points.}
\label{tab:rq2_baseline_comparison}
\footnotesize
\renewcommand{\arraystretch}{0.98}
\begin{minipage}[t]{0.35\textwidth}
\centering
\vspace{0pt}
\textbf{(a) Model-level comparison}\par\smallskip
\setlength{\tabcolsep}{1.3pt}
\renewcommand{\arraystretch}{0.98}
\begin{tabular}{@{}lcc@{}}
\toprule
\textbf{Method} & \textbf{Macro-F1 (\%)} & \textbf{$\Delta$ (pp)} \\
\midrule
\rule[-0.60em]{0pt}{2.20em}Baseline & 29.3 & -- \\
\rule[-0.60em]{0pt}{2.20em}GPT-5.5 & 33.8 & +4.5 \\
\rule[-0.60em]{0pt}{2.20em}Composer-2.5 & \textbf{51.2} & \textbf{+21.9} \\
\rule[-0.60em]{0pt}{2.20em}Opus-4.8 & 20.7 & $-8.6$ \\
\rule[-0.60em]{0pt}{2.20em}DeepSeek-V4-Pro & 25.3 & $-4.0$ \\
\midrule
\rule[-0.60em]{0pt}{2.20em}\textbf{Four-model mean} & \textbf{32.7} & \textbf{+3.4} \\
\bottomrule
\end{tabular}
\end{minipage}\hfill
\begin{minipage}[t]{0.63\textwidth}
\centering
\vspace{0pt}
\textbf{(b) Pattern-level comparison}\par\smallskip
\setlength{\tabcolsep}{3pt}
\begin{tabular}{@{}lrrrrr@{}}
\toprule
\textbf{Pattern} & \textbf{Reports} & \textbf{Prev. (\%)} & \textbf{Baseline F1} & \textbf{Mean LLM F1} & \textbf{$\Delta$ (pp)} \\
\midrule
A1 & 48  & 23.0 & 37.4 & 35.4 & $-2.0$ \\
B1 & 65  & 31.1 & 47.4 & 39.2 & $-8.2$ \\
B2 & 53  & 25.4 & 40.5 & 41.9 & +1.4 \\
B3 & 150 & 71.8 & 83.6 & 57.5 & $-26.1$ \\
C1 & 45  & 21.5 & 35.4 & 37.6 & +2.2 \\
C2 & 8   & 3.8  & 7.4  & 19.3 & +11.9 \\
C3 & 10  & 4.8  & 9.1  & 24.3 & +15.2 \\
D1 & 13  & 6.2  & 11.7 & 23.5 & +11.8 \\
D2 & 6   & 2.9  & 5.6  & 16.1 & +10.5 \\
E1 & 41  & 19.6 & 32.8 & 32.9 & +0.1 \\
E2 & 12  & 5.7  & 10.9 & 32.3 & +21.4 \\
\midrule
\textbf{Macro Avg.} & -- & -- & \textbf{29.3} & \textbf{32.7} & \textbf{+3.4} \\
\bottomrule
\end{tabular}
\end{minipage}
\end{table}

\subsection{RQ2: Effectiveness of LLMs in CVV Detection}  
To evaluate LLM-based CVV detection, we select four mainstream models: GPT-5.5, Opus-4.8, Composer-2.5, and DeepSeek-V4-Pro. The selection considers availability, code understanding, reasoning capability, and model-family diversity. We also conduct small-scale pilot experiments on GPT-5.4 and Qwen3.7-Max. Because they underperform on sampled CVV cases, we exclude them from the formal evaluation for efficiency and comparability.

\textbf{Prompt design.}
Following prior work on LLM-based smart contract vulnerability detection~\cite{LLM-detect-2}, we use a taxonomy-guided multi-label prompt that provides an auditor role, definitions of all 11 CVV patterns, the source-code input, analysis requirements, and a unified JSON schema. In one response, the model determines whether any CVV is present and, for each predicted pattern, reports the vulnerable function, code evidence, reasoning, and a brief exploit path; it also identifies any missing context. We apply the same prompt template across all models. Outputs that cannot be parsed or do not contain a valid decision are treated as negative predictions.

\textbf{Constant baseline.}
Because each audit report may contain multiple CVV patterns and their report-level frequencies are imbalanced, we include an all-pattern constant baseline that predicts all 11 CVV patterns for every report. For a pattern with report-level prevalence $p$, this baseline has precision $p$, recall $1$, and an F1 score of $2p/(1+p)$. We compute the baseline F1 for each pattern and macro-average the resulting scores across all 11 patterns. This baseline provides a prevalence-aware reference for determining whether a model offers discriminative value beyond predicting every pattern as present.

\textbf{Evaluation results.}
We evaluate the four LLMs using precision, recall, and F1 score, and report macro-averages across the 11 CVV patterns. As shown in Table~\ref{tab:rq2_llm_results}, current LLMs remain limited for CVV detection, with large performance gaps across models. Composer-2.5 performs best, achieving the highest F1 score of 51.2\% and recall of 75.6\%, suggesting more aggressive detection. In contrast, GPT-5.5 achieves the highest precision of 45.4\% but only 29.8\% recall, suggesting a conservative strategy with fewer false positives but more missed CVVs. Opus-4.8 and DeepSeek-V4-Pro obtain F1 scores of 20.7\% and 25.3\%, respectively.

As summarized in Table~\ref{tab:rq2_baseline_comparison}(a), the all-pattern constant baseline achieves a Macro-F1 of 29.3\%. Composer-2.5 exceeds it by 21.9 percentage points, while GPT-5.5 provides a smaller improvement of 4.5 points. In contrast, Opus-4.8 and DeepSeek-V4-Pro fall below the baseline by 8.6 and 4.0 points, respectively. Thus, current LLMs do not consistently demonstrate discriminative CVV-detection ability beyond a prevalence-exploiting constant predictor.

\finding{5}{LLM-based CVV detection shows limited discriminative ability: only two of four models outperform the 29.3\% Macro-F1 constant baseline.}

As shown in Table~\ref{tab:rq2_baseline_comparison}(b), B3 has the highest raw mean F1 (57.5\%). However, it appears in 71.8\% of the evaluated reports, yielding an F1 of 83.6\% for the constant baseline; only Composer-2.5 exceeds this reference. Thus, B3's high raw F1 does not imply that it is intrinsically easier to detect.

Across all 11 patterns, raw F1 is strongly correlated with report-level prevalence (Spearman's $\rho = 0.955$)~\cite{spearman1904}. Accordingly, the low raw F1 scores of D2 (16.1\%), C2 (19.3\%), and D1 (23.5\%), as well as category-level differences such as Execution Integrity (46.2\%) versus Verification Soundness (19.8\%), should be interpreted descriptively rather than as direct measures of detection difficulty.

\finding{6}{Raw per-pattern F1 is strongly confounded by label prevalence ($\rho=0.955$); for B3, despite the highest mean F1, only Composer-2.5 exceeds the prevalence-based constant baseline.}

\subsection{RQ3: Error Analysis of CVV Detection} 

To answer RQ3, we manually analyze the false negatives and false positives produced by the four LLMs, and summarize the main causes of missed and incorrect CVV reports.

\textbf{False negatives.} We identify two main causes of missed CVVs. 
\textbf{(1) Missing cross-component context.} CVVs often span gateways, messengers, validators, relayers, and application contracts. When input contains only a local contract or partial files, key verification logic may be missing, causing models to report no CVV. For example, in Axelar's deposit service audit, the unauthorized token-sending vulnerability depends on interactions among deposit addresses, \emph{DepositReceiver}, and the gateway. Given only source-side helper contracts, models classify it as safe, despite noting the missing gateway logic in \emph{missing\_context}. \textbf{(2) Insufficient fine-grained semantics.} Patterns such as C2, D1, D2, and E2 are often missed because they depend on protocol-specific rules rather than explicit missing checks. For instance, in Axelar's NCC audit, models detect session-level signature reuse (C2) but miss the ground-truth D1 instance, where weighted-multisig validity depends on signature ordering.

\finding{7}{Missing cross-component context is the primary cause of CVV false negatives, while insufficient fine-grained semantics further limits semantics-heavy pattern detection.}

\textbf{False positives.} We identify two main causes of incorrect CVV reports. \textbf{(1) Surface-level pattern matching.} Models often associate keywords or local code fragments, such as \emph{trust root}, \emph{validator set}, \emph{stale state}, and \emph{execute}, with CVV patterns without verifying whether the corresponding mechanism is actually missing or misused. For example, Cannon's fault-proof VM is misclassified as C3 because its Merkle witnesses and state roots resemble stale-state reuse, although they serve a different verification purpose. \textbf{(2) Misjudged trust boundaries.} Models may also ignore that signature verification, replay protection, and validator checks are often delegated to gateways or messaging frameworks. For example, in an Axelar application contract, the execution function does not directly check the source-chain parameter. However, this is safe because the upstream gateway has already authenticated the cross-chain message. Models miss this trust boundary and incorrectly report it as an A1 CVV.

\finding{8}{CVV false positives mainly stem from surface-level pattern matching and failures to recognize cross-component trust boundaries.}

Taken together, the false-negative and false-positive cases reveal a common limitation of LLM-based CVV detection: models struggle to reconstruct complete verification flows and reason about protocol-level semantics. As a result, they may miss vulnerabilities with cross-component evidence, falsely report checks delegated to upstream frameworks, or rely on keywords for semantics-heavy patterns involving replay protection, cryptographic verification, validator sets, and trust-root management.

\finding{9}{Overall, LLM-based CVV detection is mainly limited by incomplete cross-chain verification context and gaps in cross-chain semantic reasoning.}

%% file: section-06-Discussion.tex
\begin{table}[t]
\centering
\footnotesize
\setlength{\tabcolsep}{1.5pt}
\renewcommand{\arraystretch}{1.08}
\caption{Real-world incidents involving CVVs in 2026.}
\label{tab:real-world-cvv-incidents}
\begin{tabularx}{\columnwidth}{@{}p{2.50cm}>{\centering\arraybackslash}p{1.10cm}>{\centering\arraybackslash}p{1.60cm}>{\raggedright\arraybackslash}X >{\centering\arraybackslash}p{0.65cm}@{}}
\toprule
\textbf{Bridge} &
\textbf{Date} &
\textbf{Loss} &
\textbf{Vulnerability Cause} &
\textbf{CVV} \\
\midrule

Taiko~\cite{taiko}
& 06/21
& \$1.7M
& Compromised SGX key bypassed enclave attestation
& E2 \\

Aztec~\cite{aztec}
& 06/17
& \$2.2M
& Invalid escape-hatch proof enabled unbacked withdrawal
& D \\

Secret~\cite{secret}
& 06/10
& \$4.7M
& Missing source-channel binding in CW20--ICS20 authentication
& A1 \\

Syscoin~\cite{syscoin}
& 06/08
& \$10.0M
& Duplicate commitments caused cross-layer verification mismatch
& D \\

Alephium~\cite{alephium}
& 05/30
& \$0.8M
& Re-observation authenticated forged bridge events
& A1 \\

Butter~\cite{butter}
& 05/20
& \$0.2M
& Packed-encoding collision forged retry messages
& D2 \\

\mbox{Verus--Ethereum~\cite{verus}}
& 05/18
& \$11.6M
& Missing value binding enabled unbacked payouts
& B1 \\

Adshares~\cite{adshares}
& 05/15
& \$0.6M
& Nonexistent txIDs enabled wrapped-token minting
& D \\

Kelp DAO~\cite{kelp}
& 04/18
& \$292.0M
& Compromised single-DVN path attested forged messages
& E1 \\

Hyperbridge~\cite{hyperbridge}
& 04/13
& \$2.5M
& Out-of-bounds MMR leaf accepted as valid
& A1 \\

CrossCurve~\cite{crosscurve}
& 02/01
& \$3.0M
& Express path bypassed Axelar message validation
& A1 \\

\bottomrule
\end{tabularx}
\end{table}

\section{Discussion}
\subsection{Real-world Exploitation and Impact}
CVVs are not merely audit findings, but have been repeatedly exploited in real-world bridge attacks. 
To assess their impact, we analyze 74 public cross-chain bridge incidents reported since 2021~\cite{slowmist_bridge_hack_db}. After excluding incidents unrelated to cross-chain verification, including phishing, DNS/BGP hijacking, and unknown causes, we identify 34 involving bridge verification or execution logic; 21 (61.8\%) are CVV-related.
Table~\ref{tab:real-world-cvv-incidents} shows that the 11 most recent CVV-related incidents in 2026 caused approximately \$329.3 million in losses. They include forged messages, invalid proofs, missing source-channel checks, gateway verification bypasses, ambiguous encoding, and trust-root or prover-management flaws. This diversity shows that \textbf{CVVs span bridge implementations and verification mechanisms but share a common behavior: the destination chain accepts invalid external-chain data.} This can enable unauthorized withdrawals, unbacked asset minting, or compromised privileged operations. These observations motivate treating CVVs as a distinct vulnerability class and developing dedicated taxonomies, datasets, and detection techniques.

\begin{figure}[!t]
    \centering
    \scalebox{1}[0.92]{\includegraphics[width=0.75\textwidth]{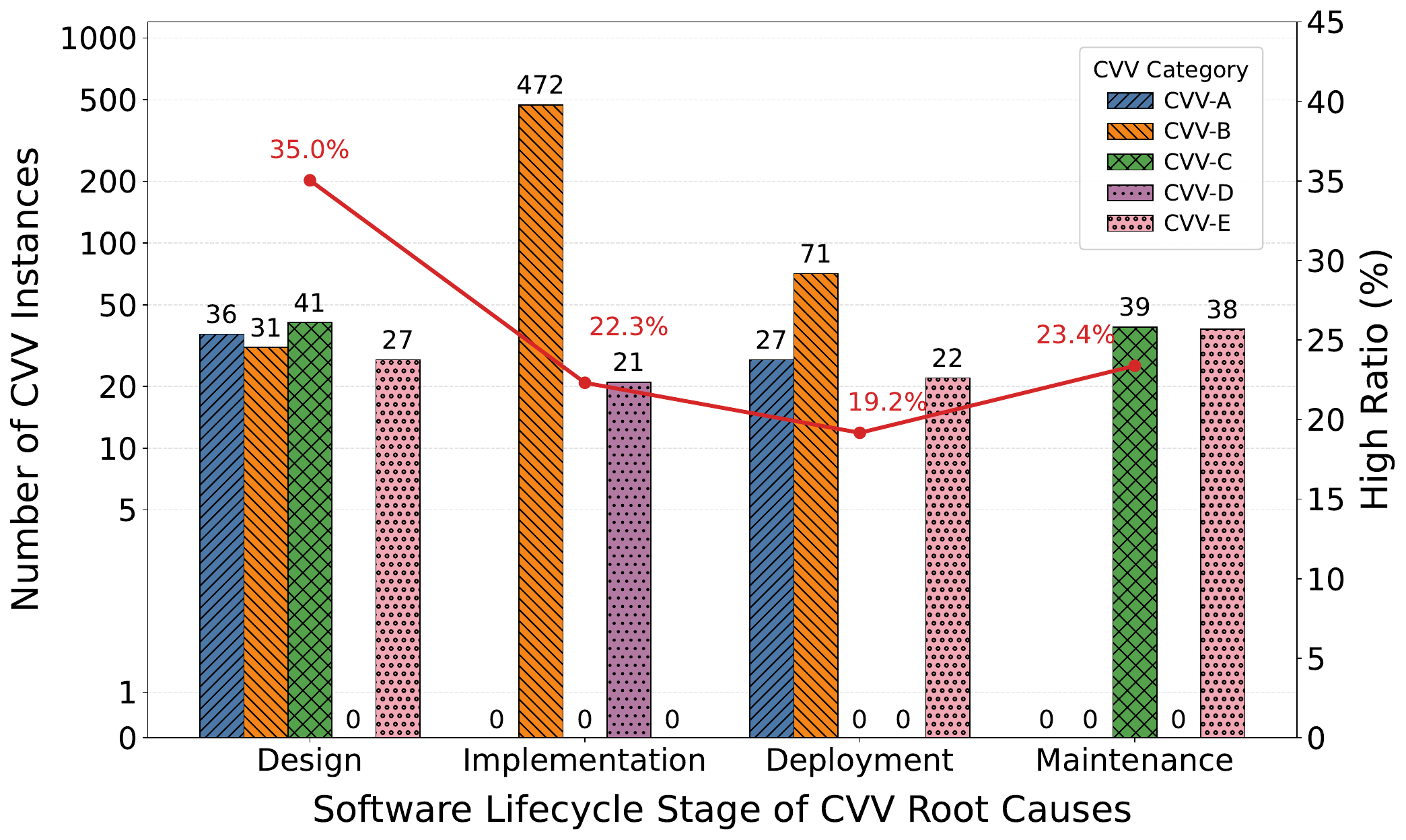}}
    \caption{Lifecycle Distribution and Severity of CVV Root Causes.}
    \Description{Grouped bars show CVV-category counts across design, implementation, deployment, and maintenance stages, together with the proportion of High-severity findings.}
    \label{fig:life_cycle}
\end{figure}

\subsection{From Design to Maintenance: Rethinking CVV Mitigation}
To examine security implications beyond individual patterns, we analyze CVVs across four lifecycle stages: design, implementation, deployment, and maintenance. We assign each CVV to its most likely origin and exclude instances whose root causes cannot be reliably determined from audit reports and code evidence.

As shown in Fig.~\ref{fig:life_cycle}, CVVs arise throughout the lifecycle. Implementation accounts for the largest share, with 493 cases (59.76\%), mostly in CVV-B, highlighting execution constraints as a common cross-chain verification challenge. However, frequency does not imply severity. Design-stage CVVs are fewer, with 135 cases (16.36\%), but have the highest High-severity ratio, suggesting that flaws in core verification assumptions, including message identity binding, replay protection, destination routing, and trust-root management, tend to be more severe. Deployment and maintenance contribute 120 (14.55\%) and 77 (9.33\%) cases, respectively, showing that CVVs also arise after implementation through misconfiguration, stale states, validator-set rotation errors, or unsafe trust-root updates. These results justify the lifecycle perspective: \textbf{CVV mitigation should address not only implementation-stage vulnerabilities, but also verification assumptions in the design stage, deployment validation, and trust-state changes during maintenance.}

\begin{figure}[!t]
    \centering
    \scalebox{1}[0.92]{\includegraphics[width=0.72\textwidth]{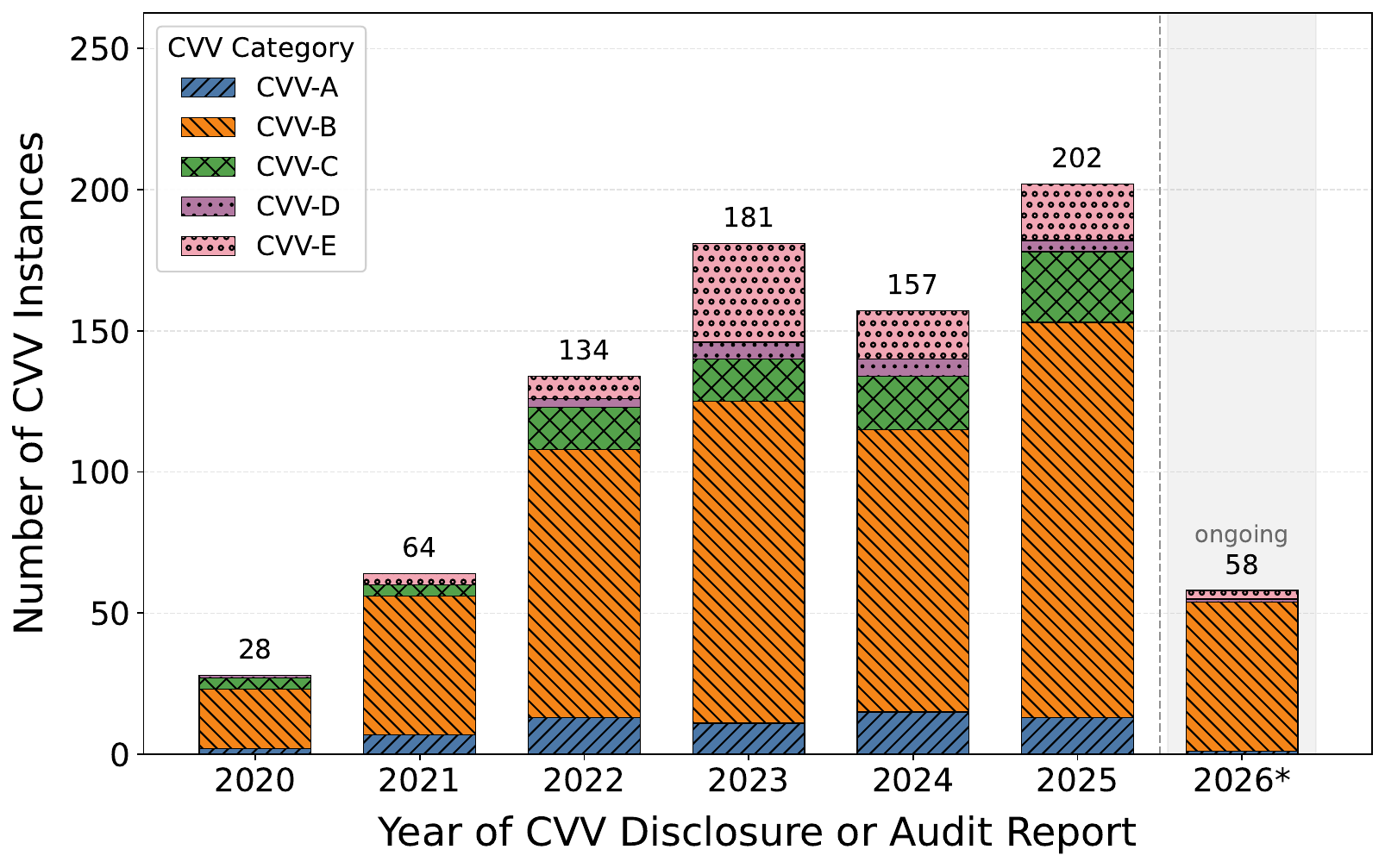}}
    \caption{Temporal Distribution of CVV Categories.}
    \Description{Stacked bars show annual counts of the five CVV categories from 2020 through the partial year 2026, with execution-integrity vulnerabilities dominating throughout.}
    \label{fig:time_trend}
\end{figure}

\subsection{How CVVs Evolve Over Time}
To examine whether CVVs are temporary vulnerabilities or persistent risks, we analyze their temporal distribution by audit-report year. As shown in Fig.~\ref{fig:time_trend}, CVV instances generally increase from 2020 to 2025, rising from 28 to 202 instances and reaching their highest level in 2025. Since 2026 is still ongoing, we treat it as a partial-year observation rather than evidence of decline. This temporal view covers 824 instances after excluding six instances with unreliable dates. The most notable trend is the persistent dominance of CVV-B, the execution-integrity category, across all years. This indicates that execution-related CVVs are not merely early-stage vulnerabilities in immature bridges. Instead, even as bridge systems evolve, contracts repeatedly fail to properly constrain cross-chain message inputs, execution parameters, destination routing, and post-verification behaviors. Therefore, \textbf{execution integrity remains a long-term structural challenge in cross-chain verification security.}

\subsection{Implications for CVV Research and Practice}
Our findings suggest four priorities for CVV research and practice. \textbf{(1) Model execution semantics beyond authentication.} Since 69.52\% of CVVs violate execution integrity, analyses should track how authenticated cross-chain data is bound to identities, destinations, parameters, and execution state. \textbf{(2) Reason across bridge components.} Because CVVs often span gateways, messengers, validators, and application contracts, detectors should model delegated checks and trust boundaries across components. \textbf{(3) Prioritize by prevalence and severity.} Less frequent patterns such as E1 and C1 still exhibit relatively high proportions of High-severity cases; triage should therefore consider both frequency and impact. \textbf{(4) Develop CVV-specific detectors.} Existing tools provide limited documented coverage, while LLMs struggle with incomplete context and protocol-specific semantics. Future detectors should combine cross-component analysis with explicit models of verification guarantees.

\subsection{Threats to Validity}
\textbf{Internal validity.} Potential biases may arise in CVV identification, taxonomy construction, code mapping, and detection evaluation. Audit reports vary in terminology, granularity, and disclosure style, which may affect manual labeling. To mitigate this threat, we define CVV inclusion criteria, classify findings by violated cross-chain verification properties, and iteratively refine the taxonomy. Code mapping may be affected by private repositories, deleted commits, pull requests, or version-diff-based targets. We cross-check repository metadata, commits, files, and functions, excluding cases with insufficient evidence from metadata-sensitive analyses. For LLM evaluation, we use the same prompt template, input format, output schema, dataset, and metrics across models. Nevertheless, outputs may vary across runs, model versions, and service configurations; our results therefore characterize the evaluated versions and settings rather than immutable model capabilities.

\textbf{External validity.} Our public-report and accessible-code dataset may underrepresent private bridges, unpublished vulnerabilities, and unaudited projects. Observed prevalence, baseline performance, and relative model performance may also depend on audit scope, auditor expertise, project maturity, and bridge architecture, and may not transfer unchanged to corpora with different pattern distributions. Future protocols or tools may further change these results. Nevertheless, our dataset covers diverse bridge projects, languages, verification mechanisms, and CVV patterns, and we release the dataset and detection results to support future research.

%% file: section-07-related-work.tex
\section{Related Work}
\subsection{Single-chain Smart Contract Vulnerability Studies}
Prior work has extensively studied single-chain smart contract vulnerabilities. Existing approaches use static analysis~\cite{Slither}, symbolic execution~\cite{Oyente}, fuzzing~\cite{ConFuzzius}, formal verification~\cite{KEVM}, and rule-based detection~\cite{SmartCheck}. They target common vulnerability patterns, including reentrancy, integer overflows, timestamp dependence, unchecked calls and unsafe delegatecall. Recent work also studies specific vulnerability classes, using empirical datasets and dedicated analyses. Representative examples include access-control vulnerabilities~\cite{AChecker}, centralization defects~\cite{Centralization-Defects}, cryptographic defects~\cite{Crypto-Fails}, proxy-contract issues~\cite{Proxy-Contracts}, and code-cloning vulnerabilities~\cite{Code-Cloning}. However, these studies are grounded in single-chain execution semantics, which rely on local state, intra-chain flows, or single-chain transactions, making them insufficient for characterizing vulnerabilities rooted in cross-chain verification semantics.

\subsection{Cross-chain Bridge Security Studies}
Recent studies have investigated bridge-specific security vulnerabilities. Xscope presents an early study of cross-chain bridge attacks and detects attack behaviors from historical transactions~\cite{Xscope}. Later studies shift from transaction-based forensics to pre-exploitation contract-level analysis. SmartAxe targets access-control incompleteness and cross-bridge semantic inconsistency in bridge contracts~\cite{smartaxe}, while XGuard detects inconsistent behaviors between source- and destination-chain contracts~\cite{XGuard}. BridgeGuard studies external interaction vulnerabilities in bridge router contracts~\cite{bridgeguard}. ChainSniper examines common vulnerabilities in cross-chain contracts~\cite{ChainSniper}. However, these works remain limited to specific scenarios, such as post-hoc forensics, bridge DApp workflows, or router-contract interactions, and do not systematically examine vulnerabilities across the cross-chain verification process.

%% file: section-08-Conclusion.tex
\section{Conclusion}

In this paper, we conduct the first systematic empirical study of cross-chain verification vulnerabilities (CVVs). By analyzing 588 audit reports from 36 real-world bridges, we derive a taxonomy of 11 CVV patterns in 5 categories and construct a dataset of 830 CVV instances. Our study characterizes their distribution, severity, and cross-project diversity, and evaluates the applicability of existing detectors and LLMs. The results show that current approaches provide limited and inconsistent CVV detection, largely due to insufficient cross-component context and cross-chain semantic reasoning. These findings highlight the need for CVV-specific analysis that models end-to-end verification semantics and for evaluation methods that account for pattern prevalence. Future work will explore such techniques to improve the security and reliability of cross-chain bridges.